\documentclass[aps,prb,reprint,floatfix,showkeys,a4paper]{revtex4-1}
\usepackage{hyperref}
\usepackage{inputenc}
\inputencoding{utf8}
\usepackage[OT1]{fontenc}
\usepackage{siunitx}
\usepackage{graphicx}
\usepackage{listings}
\usepackage{amssymb,amsmath}
\usepackage{array}
\usepackage{commath}
\usepackage{amsmath}
\usepackage{amsthm}
\usepackage{amsfonts}
\usepackage{amssymb}
\usepackage{ifthen}
\usepackage{lmodern}
\usepackage[T1]{fontenc}
\usepackage{textcomp}
\usepackage{verbatim}
\usepackage[danish,english]{babel}
\usepackage{multirow}
\usepackage{booktabs}
\usepackage{url}
\usepackage{color}
\usepackage{pspicture}
\usepackage{tikz}
\usepackage[top=2cm,bottom=2cm,left=1.5cm,right=1.5cm]{geometry}

\newcommand{\hwnote}{\color{black}}

\begin{document}
\selectlanguage{english}

\title{Connection between fragility, mean-squared
  displacement and shear modulus in two van der Waals bonded glass-forming liquids}
\author{H.W. Hansen} \email{hwase@ruc.dk} \author{B. Frick}
\author{T. Hecksher} \author{J. C. Dyre} \author{K. Niss} \affiliation{``Glass and Time'', Department of Science and Environment, Roskilde University, Postbox 260, DK-4000 Roskilde, Denmark} \affiliation{Insitut Laue-Langevin, 71 Avenue des Martyrs, F-38042 Grenoble, France}

\date{\today}

\begin{abstract}

The temperature dependence of the high-frequency shear modulus
measured in the kHz range is compared to the mean-squared
displacement measured in the nanosecond range for the two van der Waals bonded glass-forming liquids cumene and 5PPE. This provides an experimental
test for the assumption connecting two versions of the
shoving model for the non-Arrhenius temperature dependence of the
relaxation time in glass formers. The two versions of the model
are also tested directly and both are shown to work well for these
liquids. 

\end{abstract}


\maketitle 

\section{Introduction}
The glass transition happens when a
supercooled liquid falls out of equilibrium, i.e., when the structural relaxation
time is so long that the liquid cannot equilibrate within a given
experimental time. The temperature dependence of the relaxation time
in the liquid just above the glass transition is in most cases
super-Arrhenius. Liquids with a strongly super-Arrhenius behaviour are
traditionally referred to as ``fragile'' liquids following the
convention of Angell \cite{Angell85} as opposed to ``strong'' liquids
with a close to Arrhenius behaviour. It is generally assumed that the
relaxation dynamics are governed by energy barriers to be overcome by
thermal activation, similar to an activation energy for a chemical
reaction \cite{Dyre06}. In order to obtain super-Arrhenius behaviour in
this view, the activation energy, $\Delta E$ needs to be a decreasing
function of the temperature, $T$, and the relaxation time, $\tau$ is
given by
\begin{equation}\label{eq:tau_energy}
\tau(T) = \tau_0\exp\left(\frac{\Delta E(T)}{k_BT}\right),
\end{equation}
where $\tau_0\sim\SI{e-14}{\second}$ is a typical microscopic time and
$k_B$ is the Boltzmann constant. The fundamental
question is then; what causes the temperature dependence of the activation energy that almost always increases upon cooling with only a few exceptions, causing the super-Arrhenius behaviour.

In the viscous liquid just above the glass transition there is a separation of time scales between the fast thermal vibrations taking place on the order of picoseconds, and the relaxation time which has a time scale of the order of hundreds of seconds. The separation of time scales has the consequence that the liquid will appear solid-like
on time scales much shorter than the relaxation time, $\tau$, and it will show liquid behaviour on time scales much longer than $\tau$. In
the energy landscape picture \cite{Goldstein69}, this corresponds to a
separation between fast vibrations around the energy minima on short
time scales and the inherent dynamics on longer time scales,
due to jumps between potential energy minima.
 
There is no consensus on what governs the super-Arrhenius temperature
dependence of the relaxation time in liquids, though numerous models
and theories have been developed in trying to encompass the phenomenon
\cite{Dyre06,Stillinger13,Langer14,Zheng17}. The shoving model which is the
focus of this paper belongs to a class of models referred to as
elastic models \cite{Dyre96,Dyre04}.

The starting point of elastic models is that a flow event, a molecular
rearrangement, takes place on very short time scales by barrier
transition. The transition itself is a fast process, but in the
viscous liquid it is rare, which leads to slow relaxation. 
Since the transition
is fast, it is governed by properties of the liquid
at short time scales where it appears as a solid. This gives a link
between the vibrational, short-time elastic properties of the liquid
and the relaxation on long time scales. As the liquid is cooled, the
liquid hardens, the mechanical moduli increase and the vibrational
amplitudes decrease. This leads to an increase in the barrier
height which in turn leads to the super-Arrhenius behaviour of the
liquid's relaxation time. The details of the argument vary for the
different versions of the elastic models. 

There is a series of more phenomenological results, which are not
directly related to elastic models, but which support the notion
that there is a connection between fast and slow dynamics.  One of the
first was the observation in 1992 by Buchenau and Zorn of a relation
between fast and slow dynamics in
selenium \cite{Buchenau92}. They found a relation between the
temperature dependence of the slow structural relaxation, the
viscosity, and the fast mean-squared displacement (MSD) studied with
neutron time-of-flight. A connection between fast vibrational and slow
structural dynamics was also suggested in several other works (see, e.g., the references of Ref.~\onlinecite{Dyre06}). Some of these suggest a connection between the vibrational and elastic properties of the glass and the fragility of the corresponding liquid \cite{Sokolov93,Scopigno03,Novikov04}, others suggest a connection between the temperature dependence of the vibrations in the liquid and the temperature dependence of the structural relaxation time, the alpha relaxation, \cite{Larini08,Ngai04,Bernini15}, closer to the original result from Buchenau \cite{Buchenau92} and the predictions of the shoving model discussed in Sec.~\ref{sec:theory}. 

The shoving model and related elastic models have recently been
discussed in the context of several theoretical developments. In
2013 Yan, D{\"u}ring, and Wyart discussed from a general point of
view the connection between glass elasticity and fragility in a model
that connects the two properties such that elasticity is a good
predictor of fragility \cite{yan13}. Mirigian and
Schweizer proposed a unified model for the viscosity of simple liquids
going from the less-viscous regime of ``ordinary'' liquids to the
highly viscous supercooled regime, in which the deviation from
Arrhenius temperature dependence in the high-viscosity regime is
dominated by the elastic ``shoving'' work done on the surroundings to
locally lower the density \cite{mir13}. In 2015 Schirmacher, Ruocco,
and Mazzone proposed a unified theory for the viscosity, the
low-temperature alpha relaxation and the high-frequency
vibrational anomalies. The basic idea was to regard the system as a
spatial mixture of different Maxwell viscoelastic elements
characterized by a distribution of activation energies, each proportional to
the local high-frequency shear modulus \cite{sch15}. Also in 2015
Betancourt, Hanakata, Starr, and Douglas connected the short-time
vibrational MSD to free volume and cooperativity,
arguing that several apparently different models for the viscous
slowing down are, in fact, different aspects of the same mechanism \cite{bet15}. The shoving model and related elastic models have also been used recently for interpreting experimental findings, e.g., in Ref.~ \onlinecite{rou11,xu11,pot13,mir14b,kla15,kra15,mit16,ike16,syu13,Liu15}.

The shoving model exists in two different formulations, one which
connects the relaxation time to the high-frequency shear modulus,
$G_\infty$, and one which relates the relaxation time to short-time
MSD. The two versions of the model are equivalent under a few simple
approximations \cite{Dyre04}. One of these assumptions is somewhat
implicit, namely that the two properties are measured at the same time
scale -- or that they are measured in a range where there is no time
scale dependence of the properties. However, as the alpha relaxation
time becomes longer, i.e., beyond the millisecond range, many liquids
exhibit one or more beta relaxation processes at shorter time scales
than the alpha relaxation time. The beta relaxation can have a quite
large amplitude in the shear modulus \cite{Jakobsen11} and the elastic
properties and the temperature dependencies of these will therefore be
different when probed at different time scales. Many of the tests of
the $G_\infty$ version of the shoving model are made based on
measurements made on the kHz range, whereas the MSD version has been
tested primarily based on neutron scattering data performed on the
pico- or nanosecond time scale.

The issue of which time scale to use in elastic models has been
discussed previously \cite{Niss10,Buchenau14}. Since the thermal
motion that gives rise to the transition is dominated by phonons, it
is argued that the relevant time scale should be the picosecond time
scale. However, for some liquids elastic models appear to work better
when tested at longer time scales where the properties are more
temperature dependent than at the phonon-times
\cite{Buchenau14,Niss10}. In other words, the temperature dependence
of the vibration on the phonon time scale is not always large enough
to account for the super-Arrhenius temperature dependence of the
relaxation time. Based on these types of observations, Buchenau
\cite{Buchenau14} argues that the elastic models need to be combined
with an Adam-Gibbs model, and that both the hardening of the liquid
and the decrease of entropy are to be included to properly explain the
temperature dependence of the relaxation time. However, there is also
a paper where the $G_\infty$ version of the model is supported by
$G_\infty$ data determined from a range of techniques using different
time scales in order to establish the plateau value correctly
\cite{Klieber13}. In a recent review on experimental tests
\cite{Hecksher15} of both versions of the shoving model, it was found
that the shoving model works in many cases, but in other cases not,
yet there is no apparent system in when it works and when it does not
work.

{\hwnote In this paper we experimentally test the equivalence of the
  two different versions of the shoving model by comparing the
  temperature dependence of the high-frequency shear modulus to that
  of the short-time MSD. Moreover, we directly compare the performance
  of the two versions of the shoving model. To the best of our
  knowledge this is the first example of an experimental investigation
  of the assumptions made in order to arrive at the equivalence
  between the two versions of the model in
  Ref.~\onlinecite{Dyre04}. As described above the assumptions imply a
  connection between dynamics on widely different time scales and it
  is unlikely that it will work for liquids with one or more beta
  relaxations. Our aim is establish whether the assumptions can lead
  to a coherent picture that is consistent with experimental data in
  the simple case where there are no additional
  relaxations. Therefore, we turn to liquids showing as simple
  behaviour as possible.} Both liquids have been found to obey density
scaling, which means that the relaxation time is a unique function of
$\rho^\gamma/T$, where $\rho$ is density, $T$ is temperature and
$\gamma$ is a material constant
\cite{Gundermann13_thesis,Xiao15,Hansen16}. Moreover, they obey
time-temperature superposition (TTS), which means that the spectral
shape is independent of temperature
\cite{Jakobsen05,Niss07_thesis}. Shear mechanical and dielectric
spectroscopy measured on cumene (see Fig.~\ref{fig:msd_shear_cumene}
of this paper and Ref.~\onlinecite{Hansen16}) show a very low
amplitude beta relaxation (in the range of percent of the alpha
relaxation) whereas 5PPE only exhibits a weak wing
\cite{Jakobsen05}. The absence of a prominent beta relaxation should
ensure that the elastic shear modulus does not change appreciably in
the time scale from milliseconds to seconds.

The paper is structured as follows. Section~\ref{sec:theory}
introduces the two versions of the shoving model tested in this paper
and the underlying assumptions. In Sec.~\ref{sec:experiments} we
present the data of the two studied liquids. In
Sec.~\ref{sec:testofmodels}, we test the models and also present our
interpretation of the data, before discussing our findings in
Sec.~\ref{sec:discussion}.


\section{The shoving model -- two versions}\label{sec:theory}
In the original $G_\infty$ version of the shoving model
\cite{Dyre96,Dyre98,Dyre06,Hecksher15}, a local expansion is assumed
to take place in order for a flow event to happen. The activation
energy is identified as the work done \emph{shoving} aside the
surrounding liquid during this local expansion, and the activation
energy is associated with the elastic energy located in the
surroundings of the flow event. According to the shoving model,
the surrounding liquid will behave like a solid during the expansion because the flow event itself is fast. {\hwnote Assuming the local
  region that expands is spherical, the relevant elastic constant of
  the surroundings is the elastic shear modulus\cite{Dyre96},
  $G_\infty$. Moreover it can be shown that the main
  contribution to elastic energy is shear elastic energy and that the
  bulk elastic energy only plays a minor role far from an arbitrary
  point defect in an isotropic solid, no matter how large the bulk
  modulus is compared to the shear modulus \cite{Dyre07}.}

The temperature dependence of the relaxation time according to the
shoving model is given by
\begin{equation}\label{eq:tau_shear}
\tau(T)=\tau_0\exp\left(\frac{V_c G_\infty(T)}{k_BT}\right),
\end{equation} 
where $V_c$ is a characteristic molecular volume which is assumed to
be constant. 

In the MSD version of the shoving model \cite{Dyre04}, the
activation energy is related to the MSD
associated with molecular vibrations taking place on time scales
where the glass-forming liquid acts like a solid. The idea is that
larger vibrations are connected to a softer potential, which leads to a
smaller energy barrier. The MSD version of the shoving model is given
by 
\begin{equation}\label{eq:tau_msd}
\tau(T)=\tau_0\exp\left(\frac{a^2}{\langle u^2\rangle(T)}\right),
\end{equation}
where $\langle u^2\rangle(T)$ is the vibrational MSD, and $a$ is a characteristic molecular length assumed to be
constant.

The approximate equivalence between the two versions of the shoving model is derived
by modelling the vibrations harmonically and averaging over the two
types of phonons, yielding \cite{Dyre04}
\begin{equation}
\langle u^2\rangle(T)\propto T\left(\frac{2}{G_\infty(T)}+\frac{1}{M_\infty(T)}\right).
\end{equation}
where $G_\infty$ and $M_\infty$ are the transverse and longitudinal
moduli, respectively.  {\hwnote It can be shown that the temperature dependence
of the shear modulus dominates the total temperature dependence of the
expression\cite{Dyre04}, leading to}
\begin{equation}\label{eq:harm_approx}
\langle u^2\rangle(T)\propto \frac{T}{G_\infty(T)}.
\end{equation} 

Combining Eq.~(\ref{eq:harm_approx}) with Eq.~(\ref{eq:tau_msd}) gives the equivalence of the two versions of the
shoving model and one ends up with three proportional terms:
\begin{equation}\label{eq:harmapprox}
\frac{\Delta E(T)}{k_BT}\propto\frac{V_cG_\infty(T)}{k_BT}\propto\frac{a^2}{\langle u^2\rangle(T)}.
\end{equation}


\section{The Experiments and Data}\label{sec:experiments}
We present new MSD and inelastic fixed window scans \cite{Frick12} measured with neutron backscattering
as well as new data on the shear modulus measured by broadband
shear-mechanical spectroscopy on the liquid  cumene (isopropyl benzene). Cumene has been studied for many years with other
techniques, for example in
Refs.~\cite{Bridgman49,Barlow66,Li95,Masahara06,Niss10}. Cumene is a fragile liquid ($m\approx70$) with only a very
small beta relaxation. 

In Sec.~\ref{sec:5PPE} we also test the elastic models for another van
der Waals bonding liquid, a 5-polyphenyl ether (5PPE), which is a large
molecule but with behaviour and fragility similar to that of cumene
\cite{Xiao15,Roed15}. For 5PPE we present new MSD data and compare to
earlier published shear mechanical data.

Cumene was purchased from Sigma Aldrich, and 5PPE was purchased from Santolubes. Both were used as acquired.

\subsection{Mean-squared displacement}
The MSD is measured by elastic incoherent
neutron scattering. Elastic temperature scans at the backscattering
instrument IN16B were performed for this study at the Institut
Laue-Langevin (ILL).

Neutron backscattering can be used to study fast dynamics of atoms by
measuring the incoherent intermediate scattering function,
$I(Q,t)$. The incoherent intermediate scattering function is the space Fourier
transform of the density self-correlation function, which gives the
probability that an atom at some time, $t$, is at a given position at a
new time, $t+t'$.

The elastic scans were performed with an energy resolution of $\Delta
E\approx\SI{0.75}{\micro\eV}$, accessing a time scale of the dynamics
of around $\SI{5}{\nano\second}$. The energy resolution of the
instrument corresponds to studying the dynamics at a specific time,
$t$. In elastic scans, $I(Q,t)$ is essentially time independent and
only dependent on the scattering vector, $Q$ and the temperature,
$T$. We therefore introduce the incoherent intermediate scattering
function notation $I(Q,T)$ used in incoherent elastic neutron
scattering.

The MSD is obtained from the data using the
Gaussian approximation \cite{Rahman62}
\begin{equation}\label{eq:gaussianapprox}
I(Q,T)=\exp\left(\frac{-Q^2\langle u^2\rangle(T)}{3}\right),
\end{equation}
which is valid if the distributions of displacements is Gaussian, for
example in the case for harmonic vibrations.

\begin{figure}[htbp!]
\centering
\includegraphics[width=0.6\columnwidth]{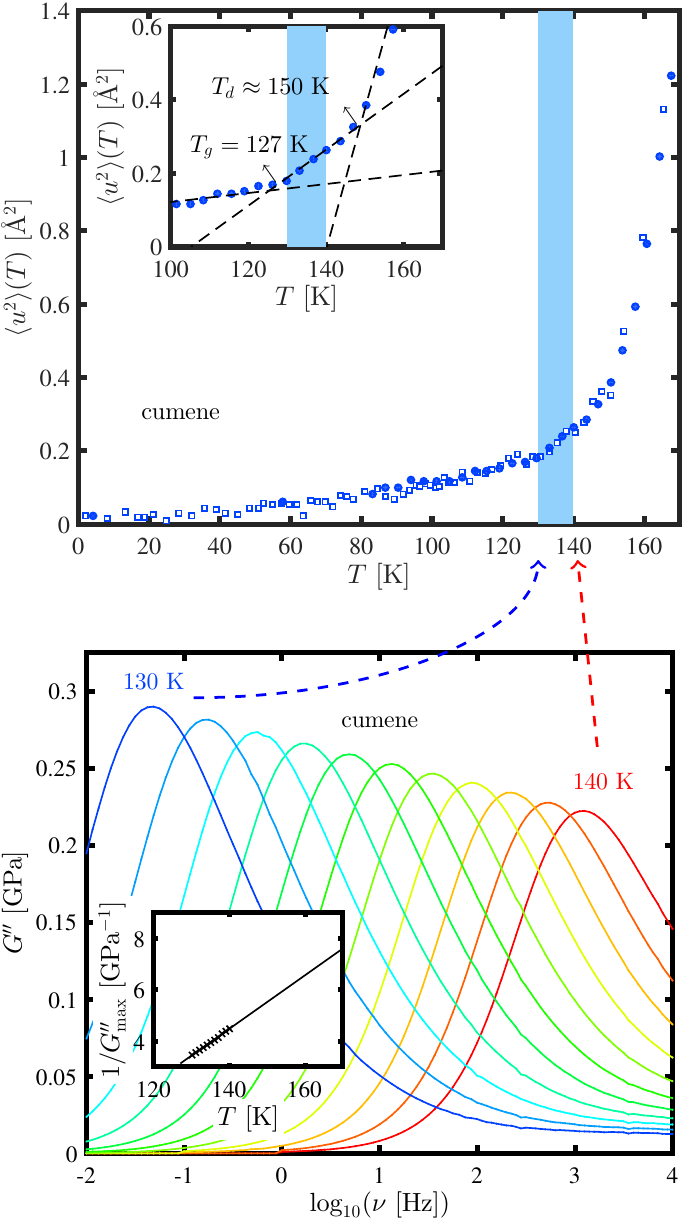}
\caption{Top: The MSD of cumene as a function of temperature from IN16B ($\bullet$) and old data from IN10 ({\tiny $\Box$}). The shaded area marks the temperature interval where the shear modulus was measured. Inset: The lines are guides to the eye to show the change in dynamics around $T_g$ and $T_d\approx\SI{150}{\kelvin}$. 
Bottom: Loss peak of the shear modulus of cumene measured in the temperature interval $130-\SI{140}{\kelvin}$. The inset shows the extrapolation of the loss-peak moduli according to Eq.~(\ref{eq:shear_extrap}) into the higher-temperature liquid range that was used for neutron scattering.}\label{fig:msd_shear_cumene}
\end{figure}

The MSD is calculated from the logarithm of the elastic intensity for
each temperature as a function of $Q^2$ according to
Eq.~(\ref{eq:gaussianapprox}). The data for each temperature is
normalized to the data at the lowest temperature, $T=\SI{5}{\kelvin}$,
thus removing any zero-point motion. The MSD of cumene as a function
of temperature is shown in Fig.~\ref{fig:msd_shear_cumene}. Around the glass transition ($\tau_\alpha=\SI{100}{s}$) for cumene\cite{Hansen16} at $T_g=\SI{127}{\kelvin}$ there is a change in slope of the
MSD as a function of temperature. We see a collapse of previous data
measured on IN10 at ILL \cite{Niss10} with the new data with better
statistics from IN16B.

\subsection{Shear modulus}
The shear modulus was measured as a function of frequency using a
piezo-ceramic transducer \cite{Christensen95} in the frequency
interval $10^{-2}-\SI{e4}{\hertz}$. The loss peak of the shear modulus
for cumene is shown in Fig.~\ref{fig:msd_shear_cumene} in the
temperature interval $130-\SI{140}{\kelvin}$ probed in steps of
$\SI{1}{\kelvin}$. This temperature range corresponds to the
shaded area in the MSD plot
(Fig.~\ref{fig:msd_shear_cumene}). Note that within a temperature range of
\SI{10}{\kelvin}, the alpha relaxation time changes roughly five orders of
magnitude.

The inverse of the frequency of the shear loss peak maximum,
$\nu_\mathrm{max}$, gives a measure of the alpha relaxation time,
$\tau_\alpha=1/(2\pi\nu_\mathrm{max})$. For studying the shoving model, we also need the elastic shear modulus
(Eqs.~(\ref{eq:harmapprox})~and~(\ref{eq:tau_shear})). To establish whether
a plateau in the real part of the shear modulus is actually reached
is not easy \cite{Dyre12}, especially for
higher temperatures within the frequency range of this setup. However, if a liquid obeys TTS, {\hwnote i.e., the spectral shape does not change with time and temperature, given the Kramers-Kronig relations between the real and imaginary part of the shear modulus}, the plateau of the elastic shear modulus is proportional to the maximum loss, $G_\infty(T)\propto G''_\mathrm{max}(T)$, i.e., they have the same temperature dependence \cite{Hecksher15}.

Cumene obeys TTS with only a very small beta relaxation. Since the maximum of the loss
peak is more readily accessible than the
elastic (plateau) shear modulus, we will use the maximum of the loss 
in studying the elastic models throughout this paper.

The inset in Fig.~\ref{fig:msd_shear_cumene} shows the extrapolation in
temperature of the maximum shear loss for the entire liquid temperature
range used in neutron scattering, i.e., up to \SI{170}{\kelvin}. The relation from Barlow \emph{et
al.} \cite{Barlow67,Harrison76} is used to extrapolate to higher
temperatures:
\begin{equation}\label{eq:shear_extrap}
\frac{1}{G_\infty}=\frac{1}{G_0}+C(T-T_0),
\end{equation}
where $C$ is a constant. We will substitute $G''_\mathrm{max}$ for
$G_\infty$. The extrapolation is used for testing Eq.~(\ref{eq:harm_approx}) in the temperature range of the MSD in the liquid, i.e., above $T_g$.

\section{Testing the models}\label{sec:testofmodels}
\subsection{The connection between $G_\infty$ and MSD}
\begin{figure}[tp!]
\centering
\includegraphics[width=0.8\columnwidth]{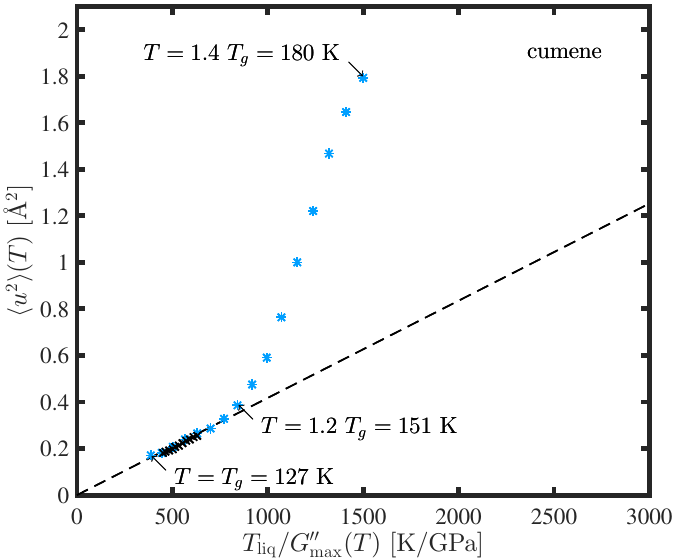} 
\caption{Testing Eq.~(\ref{eq:harm_approx}) for cumene in the liquid. The black data points mark temperatures at which the shear modulus was measured. Equation~(\ref{eq:harm_approx}) holds until the temperature $1.2~T_g$ where the alpha relaxation enters the neutron scattering window.}\label{fig:harmapprox_cumene}
\end{figure}

To test Eq.~(\ref{eq:harm_approx}), the
MSD of cumene is plotted against the shear
modulus scaled with temperature in
Fig.~\ref{fig:harmapprox_cumene}. The black data points are in the
interval where the shear data was actually measured, the rest is
the extrapolation in temperature according to
Eq.~(\ref{eq:shear_extrap}).  

The straight line is a one parameter fit, in which only the slope of
the line is fitted to the part of the data that clearly falls on a
straight line. The line shows that the data follows the
proportionality predicted by Eq.~(\ref{eq:harm_approx}). This equation
is valid under the assumption that the elastic constants measured in
the kHz range agree with the elastic constants governing the MSD
measured at roughly five orders of magnitude shorter times.  The
proportionality applies up until $1.2~T_g$. Our interpretation is that
the alpha relaxation here enters the window of the neutron scattering
instrument, causing a larger temperature dependence of the MSD than of
the shear modulus, and that the MSD grows faster than predicted from
the decrease of the shear modulus. 

\begin{figure}[htbp!]
\centering
\includegraphics[width=0.8\columnwidth]{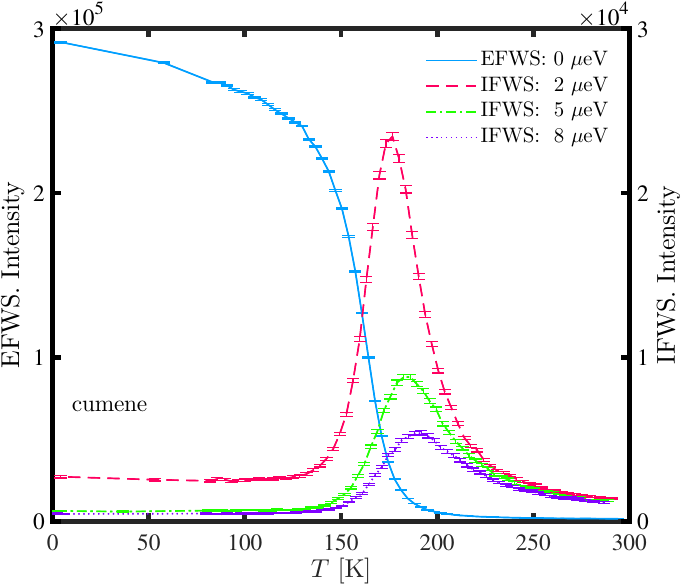}
\caption{Fixed window scan on IN16B on cumene summed over $Q$. From
  the inelastic signal (IFWS: broken lines) we see the alpha
  relaxation entering the instrument window around \SI{150}{\kelvin}
  causing a further increase in the elastic signal (EFWS: full line).
  Please note the different scales between the EFWS and
  IFWS. }\label{fig:fws_cumene}
\end{figure}

We see when the signal goes from being just elastic to also having an
inelastic contribution by use of the fixed window scan (FWS) technique
\cite{Frick12} available on IN16B at ILL. From this technique it is
possible in, for example, a temperature scan to not only gain
information about the change in elastic intensity, but also from the
inelastic intensity by continuously changing the instrument
settings. The change in elastic intensity (EFWS) and the inelastic
intensity (IFWS) for three different settings, $\Delta E = 2,5$ and
\SI{8}{\micro\eV} are shown in Fig.~\ref{fig:fws_cumene} summed over
$Q$ and normalized to monitor.

The increase of the IFWS is a sign of the alpha relaxation entering
the \SI{2}{\micro\eV} window, i.e. that it takes place at the
nanosecond time scale. This happens at the temperature
\SI{150}{\kelvin} where the relation from Eq.~(\ref{eq:harm_approx})
breaks down, causing a further increase in the elastic intensity
(Fig.~\ref{fig:harmapprox_cumene}). This is also visible in the MSD
(Fig.~\ref{fig:msd_shear_cumene}, inset of the top panel) where
another change in slope in addition to the one at the glass transition
can be seen at roughly \SI{150}{\kelvin}. This change in slope signals a dynamic transition, $T_d$, where the relaxation time and the resolution
time of an instrument intersect. This onset of dynamics was also
reported in Ref.~\onlinecite{Capaccioli12}.

\subsection{Shoving model}
%
%
Assuming that the characteristic volume $V_c$ is constant in
temperature, the shoving model predicts that the logarithm of the
relaxation time is a linear function of $G_\infty(T)/T$. The
prefactor, $\tau_0$, is given by a typical microscopic time scale. We
set it to $\tau_0=\SI{e-14}{\second}$ and define the glass transition
temperature by $\tau_g=\SI{100}{\second}$. By doing
this the linearity becomes \cite{Hecksher15} (with all times in seconds)
\begin{equation}\label{eq:shoving_predic}
\begin{split}
\log_{10}\tau(T)= (\log_{10}\tau_g-\log_{10}\tau_0)\frac{G_\infty(T)T_g}{G_\infty(T_g)T}+\log_{10}\tau_0 \\
= 16 \frac{G_\infty(T)T_g}{G_\infty(T_g)T}-14.
\end{split}
\end{equation}
{\hwnote which under the assumption $G_\infty(T)\propto G''_\mathrm{max}(T)$
introduced in the previous section yields: 
\begin{equation}
  \label{eq:1}
  \log_{10}\tau(T) = 16 \frac{G''_\mathrm{max}(T)T_g}{G''_\mathrm{max}(T_g)T}-14.
\end{equation}}

This gives rise to a ``shoving plot''; a way of testing the shoving model 
without free parameters by comparison of normalized data to the
shoving model prediction.

A similar equation can be written up for the MSD version of the
shoving model. 
\begin{equation}\label{eq:MSDshoving_predic}
\log_{10}\tau(T)=16 \frac{\langle u^2\rangle_g}{\langle u^2\rangle(T)}-14.
\end{equation}

Since the shoving model relates the relaxation time to the
short-time liquid properties, the
model is only tested in the temperature range of the
shear measurements, i.e., from $130-\SI{140}{\K}$ where we have the alpha relaxation in the frequency window of the shear modulus.
\begin{figure}[htbp!]
\centering
\includegraphics[width=0.8\columnwidth]{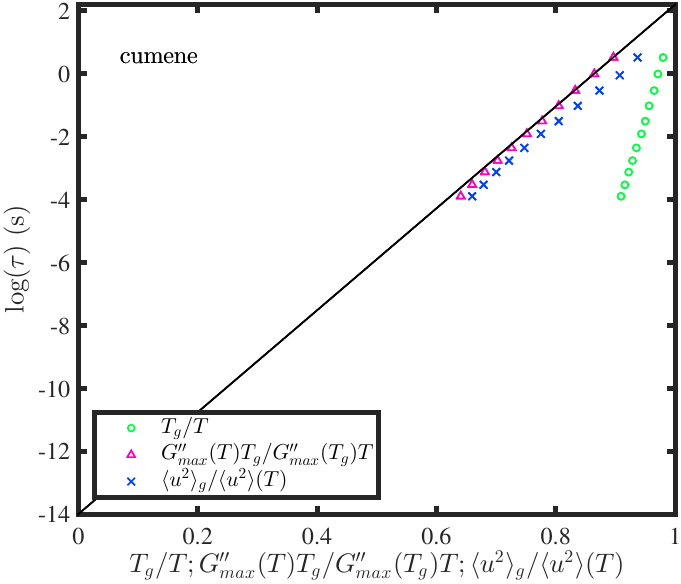}
\caption{The shoving plot with the prediction (black line), relaxation time against $\frac{G''_\mathrm{max}(T)T_g}{G''_\mathrm{max}(T_g)T}$ ({\tiny $\bigtriangleup$}), and $\langle u^2\rangle_g/\langle u^2\rangle(T)$ ({\tiny $\times$}). Relaxation time for cumene is plotted against temperature for the standard Angell plot ($\circ$).}\label{fig:shoving_cumene}
\end{figure}
In Fig.~\ref{fig:shoving_cumene}, the shoving plot with the black line
as the prediction (Eq.~(\ref{eq:shoving_predic})) is plotted along
with the scaled shear modulus, the parameter
$\frac{G_\infty(T)T_g}{G_\infty(T_g)T}$ from
Eq.~(\ref{eq:shoving_predic}), and along with the MSD scaled to the
MSD at the glass transition temperature
(Eq.~(\ref{eq:MSDshoving_predic})). The MSD data points were
interpolated to find the MSD at the temperatures where the shear
modulus was measured.  The prediction that the short-time dynamics
scales linearly with the logarithm of the relaxation time all the way
from the glass transition ($\tau=\SI{100}{\s}$) to microscopic time
scales ($\tau_0=\SI{e-14}{\s}$) agrees with the data. {\hwnote Thus the figure
shows that both the MSD and the shear modulus version of the the
shoving model can account for the non-Arrhenius behaviour in the
temperature range studied.}

\subsection{Testing for another liquid}\label{sec:5PPE}
We also tested the elastic models for the glass-forming liquid 5PPE
(5-polyphenyl ether). 5PPE has fragility of $m\approx80$, similar to
that of cumene, and it has a similar behaviour with only a weak wing \cite{Hecksher13}. The glass transition temperature of 5PPE
from shear modulus is \SI{243}{\K}. 5PPE has been shown to obey TTS
\cite{Roed13} and is found to have very simple behaviour in the sense
defined by isomorph theory \cite{Xiao15,Roed15}.

\begin{figure}[htbp!]
\centering
\includegraphics[width=0.6\columnwidth]{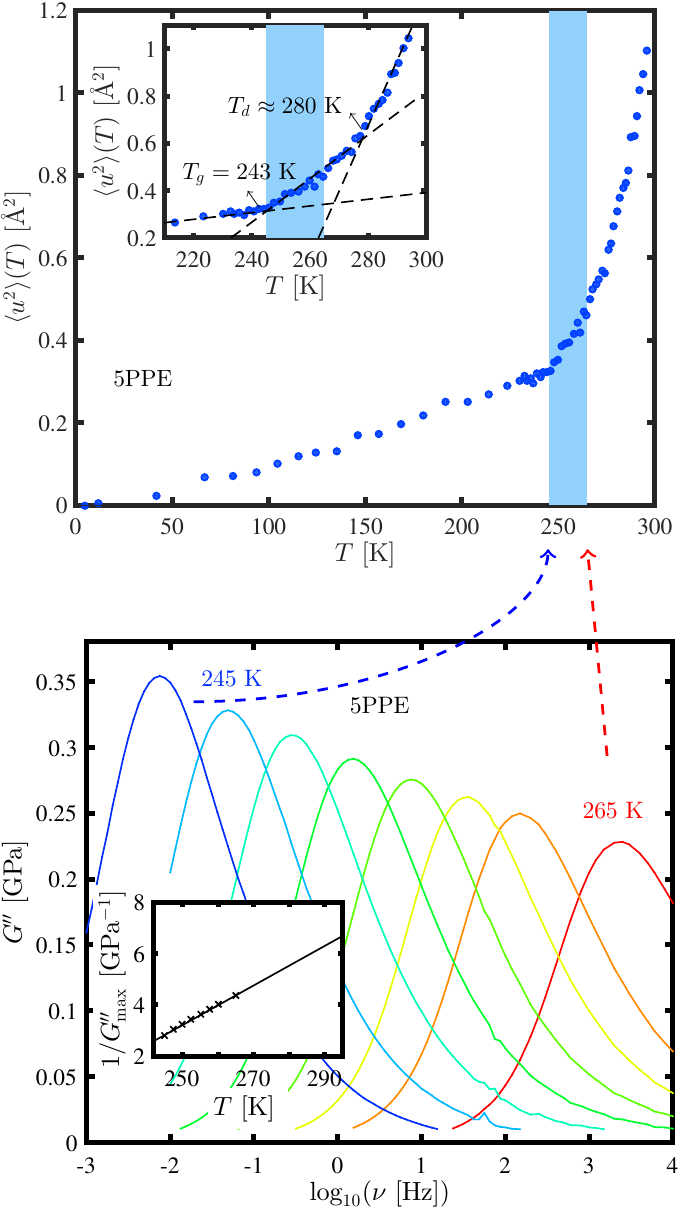}
\caption{Top: The MSD of 5PPE as function of temperature. The shaded area illustrates the temperature interval of the measured shear modulus. Inset: Zoom, the lines are guides to the eye to show the change in dynamics around $T_g$ and $T_d\approx\SI{280}{\kelvin}$. The black lines are guides to the eye.
Bottom: Loss peak of the shear modulus of 5PPE measured in the temperature interval $245-\SI{265}{\kelvin}$. Inset shows the extrapolation of the loss peak moduli into the whole liquid temperature range that was used for neutron scattering.}\label{fig:msd_shear_5PPE}
\end{figure}

The MSD shown in Fig.~\ref{fig:msd_shear_5PPE} was measured at IN16 at ILL. The shear data is from Hecksher \emph{et al.} (2013) \cite{Hecksher13} and the shear loss peaks in the temperature interval $T=245-\SI{265}{\K}$ are
also shown in Fig.~\ref{fig:msd_shear_5PPE}. The temperature range of
the shear data is marked in the MSD plot as the
shaded area. The inset shows the extrapolation into the higher
temperature region according to Eq.~(\ref{eq:shear_extrap}).

\begin{figure}[htbp!]
\centering 
\includegraphics[width=0.6\columnwidth]{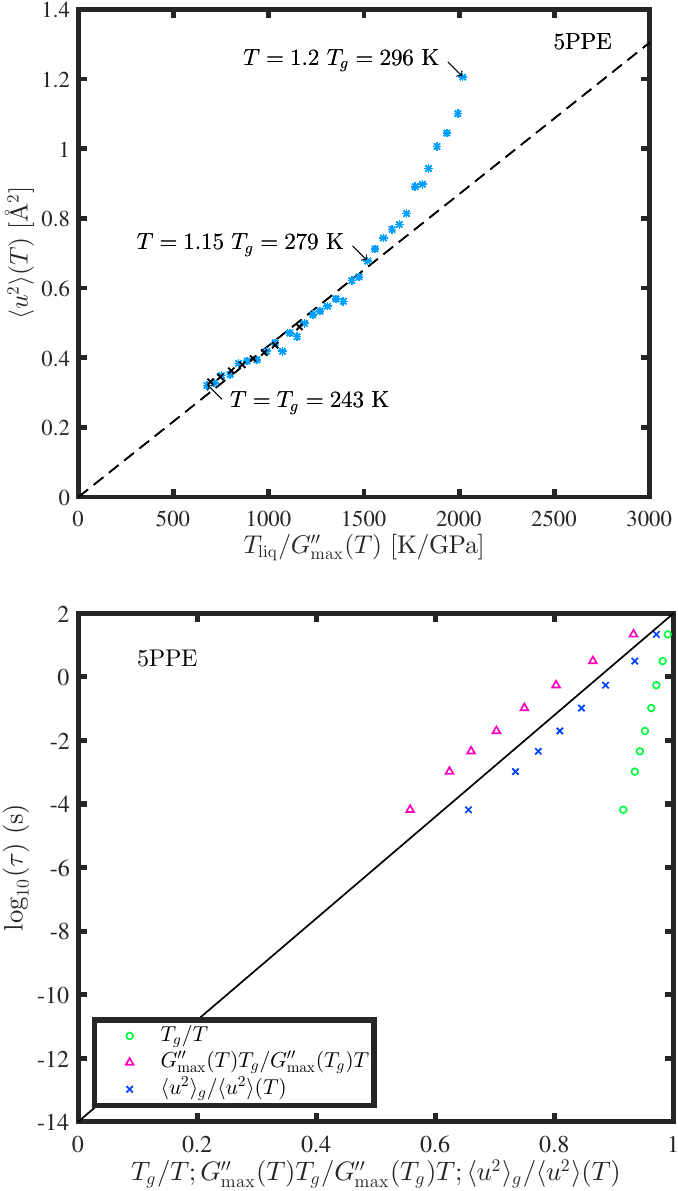}
\caption{Top: Testing Eq.~(\ref{eq:harm_approx}) for 5PPE in the liquid. The black data points mark temperatures at which the shear modulus was measured. Eq.~(\ref{eq:harm_approx}) holds until $1.15~T_g$ where the alpha relaxation enters the neutron scattering window. Bottom: The shoving plot with the prediction (black line), relaxation time against $\frac{G_\infty(T)T_g}{G_\infty(T_g)T}$ ({\tiny $\bigtriangleup$}), and $\langle u^2\rangle_g/\langle u^2\rangle(T)$ ({\tiny $\times$}). Relaxation time for 5PPE is plotted against temperature for the standard Angell plot ($\circ$).}\label{fig:elasticmodels_5PPE}
\end{figure}

The proportionality between MSD and $T/G_\infty(T)$, as well as the
shoving plot, are shown in Fig.~\ref{fig:elasticmodels_5PPE}. The
picture is the same as for cumene
(Figs.~\ref{fig:harmapprox_cumene}~and~\ref{fig:shoving_cumene}): the
elastic models work well. Regarding the proportionality between MSD
and $T/G_\infty(T)$, we see the alpha relaxation entering the neutron
scattering window at $1.15~T_g$, a slightly lower temperature than for
cumene. This could be due to the higher fragility of 5PPE. The data
follows the shoving prediction well. Clearly the scaled shear modulus
and the MSD follow the general trend predicted by the shoving model,
although not as nicely as for cumene.

\section{Discussion and Conclusion}\label{sec:discussion}
We have shown that the shoving model is confirmed in the case of the two
liquids studied, and that there is good agreement
between the two versions of the shoving model; one connecting the slow
structural relaxation to the short-time elastic modulus and one
connecting the slow structural relaxation to the short-time MSD.
This correspondence holds even though the MSD and the shear modulus are measured at two
very different time scales; the nanosecond and the millisecond,
respectively.   

For cumene, the relation between $G''_\mathrm{max}$ and the MSD shows
proportionality up to the temperature where the alpha relaxation as
seen by IFWS enters the neutron scattering window, causing a stronger
temperature dependence of the MSD than of the elastic modulus. In the
case of 5PPE we do not have the IFWS data, but we see a similar
development in the MSD. This supports the scenario proposed by
Capaccioli \emph{et al.} \cite{Capaccioli12} referring to two
transitions in the MSD of solvated proteins; the glass transition and
the dynamic transition where the relaxation time and the resolution
time of an instrument intersect.

In our view, the change in slope of the MSD at the glass transition
temperature is not due to a change in the mechanism of the nanosecond
dynamics. The dynamics is still vibrational. Rather, the modulus
 becomes much more temperature dependent because of going from
the glassy to the liquid state. {\hwnote With the assumptions we have used,} the temperature dependence just above
$T_g$ can be predicted by the change in the high-frequency
modulus. The second change is in the liquid at $T_d$ where the alpha
relaxation enters the instrument window causing a further increase in
the MSD.

{\hwnote Because of the energy-resolution dependence of the MSD, the
  study may be performed in addition on instruments with coarser
  energy resolution which also does allow to discriminate different
  vibrational and relaxational contributions to the MSD
  \cite{Buchenau92,Niss10,Frick88}. Here we present with the inelastic fixed window
  technique for the first time an alternative possibility for
  attempting a separation of the different motional contributions to
  the MSD or at least the to determine the temperature range where
  relaxation becomes important.}

For other systems with more complex behaviour such as large beta
relaxations, it is likely that the picture is more complicated. Here
we would expect a discrepancy between the temperature dependence of
the MSD at the nanosecond and the modulus measured in the
kHz range. Based on the shoving model, one expects the properties at
short time scales to be the best predictor of the temperature
dependence of the alpha relaxation time. However, literature findings
do not always support this prediction \cite{Niss10,Buchenau14}.
Another possibility is that the MSD at the
nanosecond time scale has a larger relaxational component in liquids
with a more complex relaxation map. If this fast relaxation has a
weak temperature dependence, it could lead to a relatively weaker
temperature dependence of the MSD as compared to the activation energy
and thus a deviation from Eq.~(\ref{eq:harmapprox}). Finally it is
also possible, as suggested by Buchenau \cite{Buchenau14}, that the
elastic models do not explain the full temperature dependence of the
activation energy in the general case.  In the future it is therefore
important to test the different versions of the shoving model with
liquids of different behaviour, including variations in fragility.

\newpage


\begin{thebibliography}{52}%
\makeatletter
\providecommand \@ifxundefined [1]{%
 \@ifx{#1\undefined}
}%
\providecommand \@ifnum [1]{%
 \ifnum #1\expandafter \@firstoftwo
 \else \expandafter \@secondoftwo
 \fi
}%
\providecommand \@ifx [1]{%
 \ifx #1\expandafter \@firstoftwo
 \else \expandafter \@secondoftwo
 \fi
}%
\providecommand \natexlab [1]{#1}%
\providecommand \enquote  [1]{``#1''}%
\providecommand \bibnamefont  [1]{#1}%
\providecommand \bibfnamefont [1]{#1}%
\providecommand \citenamefont [1]{#1}%
\providecommand \href@noop [0]{\@secondoftwo}%
\providecommand \href [0]{\begingroup \@sanitize@url \@href}%
\providecommand \@href[1]{\@@startlink{#1}\@@href}%
\providecommand \@@href[1]{\endgroup#1\@@endlink}%
\providecommand \@sanitize@url [0]{\catcode `\\12\catcode `\$12\catcode
  `\&12\catcode `\#12\catcode `\^12\catcode `\_12\catcode `\%12\relax}%
\providecommand \@@startlink[1]{}%
\providecommand \@@endlink[0]{}%
\providecommand \url  [0]{\begingroup\@sanitize@url \@url }%
\providecommand \@url [1]{\endgroup\@href {#1}{\urlprefix }}%
\providecommand \urlprefix  [0]{URL }%
\providecommand \Eprint [0]{\href }%
\providecommand \doibase [0]{http://dx.doi.org/}%
\providecommand \selectlanguage [0]{\@gobble}%
\providecommand \bibinfo  [0]{\@secondoftwo}%
\providecommand \bibfield  [0]{\@secondoftwo}%
\providecommand \translation [1]{[#1]}%
\providecommand \BibitemOpen [0]{}%
\providecommand \bibitemStop [0]{}%
\providecommand \bibitemNoStop [0]{.\EOS\space}%
\providecommand \EOS [0]{\spacefactor3000\relax}%
\providecommand \BibitemShut  [1]{\csname bibitem#1\endcsname}%
\let\auto@bib@innerbib\@empty
\bibitem [{\citenamefont {Angell}(1985)}]{Angell85}%
  \BibitemOpen
  \bibfield  {author} {\bibinfo {author} {\bibfnamefont {C.~A.}\ \bibnamefont
  {Angell}},\ }in\ \href@noop {} {\emph {\bibinfo {booktitle} {Relaxations in
  complex systems}}},\ \bibinfo {editor} {edited by\ \bibinfo {editor}
  {\bibfnamefont {K.}~\bibnamefont {Ngai}}\ and\ \bibinfo {editor}
  {\bibfnamefont {G.}~\bibnamefont {Wright}}}\ (\bibinfo  {publisher} {US Dpt
  of Commerce},\ \bibinfo {year} {1985})\BibitemShut {NoStop}%
\bibitem [{\citenamefont {Dyre}(2006)}]{Dyre06}%
  \BibitemOpen
  \bibfield  {author} {\bibinfo {author} {\bibfnamefont {J.~C.}\ \bibnamefont
  {Dyre}},\ }\href {\doibase 10.1103/RevModPhys.78.953} {\bibfield  {journal}
  {\bibinfo  {journal} {Reviews of Modern Physics}\ }\textbf {\bibinfo {volume}
  {78}},\ \bibinfo {pages} {953} (\bibinfo {year} {2006})}\BibitemShut
  {NoStop}%
\bibitem [{\citenamefont {Goldstein}(1969)}]{Goldstein69}%
  \BibitemOpen
  \bibfield  {author} {\bibinfo {author} {\bibfnamefont {M.}~\bibnamefont
  {Goldstein}},\ }\href {\doibase 10.1063/1.1672587} {\bibfield  {journal}
  {\bibinfo  {journal} {Journal of Chemical Physics}\ }\textbf {\bibinfo
  {volume} {51}},\ \bibinfo {pages} {3728} (\bibinfo {year}
  {1969})}\BibitemShut {NoStop}%
\bibitem [{\citenamefont {Stillinger}\ and\ \citenamefont
  {Debenedetti}(2013)}]{Stillinger13}%
  \BibitemOpen
  \bibfield  {author} {\bibinfo {author} {\bibfnamefont {F.~H.}\ \bibnamefont
  {Stillinger}}\ and\ \bibinfo {author} {\bibfnamefont {P.~G.}\ \bibnamefont
  {Debenedetti}},\ }\href {\doibase 10.1146/annurev-conmatphys-030212-184329}
  {\bibfield  {journal} {\bibinfo  {journal} {Annual Review of Condensed Matter
  Physics}\ }\textbf {\bibinfo {volume} {4}},\ \bibinfo {pages} {263} (\bibinfo
  {year} {2013})}\BibitemShut {NoStop}%
\bibitem [{\citenamefont {Langer}(2014)}]{Langer14}%
  \BibitemOpen
  \bibfield  {author} {\bibinfo {author} {\bibfnamefont {J.~S.}\ \bibnamefont
  {Langer}},\ }\href {http://stacks.iop.org/0034-4885/77/i=4/a=042501}
  {\bibfield  {journal} {\bibinfo  {journal} {Reports on Progress in Physics}\
  }\textbf {\bibinfo {volume} {77}},\ \bibinfo {pages} {042501} (\bibinfo
  {year} {2014})}\BibitemShut {NoStop}%
\bibitem [{\citenamefont {Zheng}\ and\ \citenamefont {Mauro}(2017)}]{Zheng17}%
  \BibitemOpen
  \bibfield  {author} {\bibinfo {author} {\bibfnamefont {Q.}\ \bibnamefont
  {Zheng}}\ and\ \bibinfo {author} {\bibfnamefont {J.~C.}\ \bibnamefont
  {Mauro}},\ }\href {\doibase 10.1111/jace.14678} {\bibfield  {journal}
  {\bibinfo  {journal} {Journal of American Ceramic Society}\ }\textbf {\bibinfo {volume} {100}},\ \bibinfo {pages}
  {6--25} (\bibinfo {year} {2017})}\BibitemShut {NoStop}%
%
%
\bibitem [{\citenamefont {Dyre}\ \emph {et~al.}(1996)\citenamefont {Dyre},
  \citenamefont {Olsen},\ and\ \citenamefont {Christensen}}]{Dyre96}%
  \BibitemOpen
  \bibfield  {author} {\bibinfo {author} {\bibfnamefont {J.~C.}\ \bibnamefont
  {Dyre}}, \bibinfo {author} {\bibfnamefont {N.~B.}\ \bibnamefont {Olsen}}, \
  and\ \bibinfo {author} {\bibfnamefont {T.}~\bibnamefont {Christensen}},\
  }\href {\doibase 10.1103/PhysRevB.53.2171} {\bibfield  {journal} {\bibinfo
  {journal} {Physical Review B}\ }\textbf {\bibinfo {volume} {53}},\ \bibinfo
  {pages} {2171} (\bibinfo {year} {1996})}\BibitemShut {NoStop}%
\bibitem [{\citenamefont {Dyre}\ and\ \citenamefont {Olsen}(2004)}]{Dyre04}%
  \BibitemOpen
  \bibfield  {author} {\bibinfo {author} {\bibfnamefont {J.~C.}\ \bibnamefont
  {Dyre}}\ and\ \bibinfo {author} {\bibfnamefont {N.~B.}\ \bibnamefont
  {Olsen}},\ }\href {\doibase 10.1103/PhysRevE.69.042501} {\bibfield  {journal}
  {\bibinfo  {journal} {Physical Review E}\ }\textbf {\bibinfo {volume} {69}},\ \bibinfo {pages}
  {042501} (\bibinfo {year} {2004})}\BibitemShut {NoStop}%
\bibitem [{\citenamefont {Dyre}(2007)}]{Dyre07}%
  \BibitemOpen
  \bibfield  {author} {\bibinfo {author} {\bibfnamefont {J.~C.}\ \bibnamefont
  {Dyre}},\ }\href {\doibase 10.1103/PhysRevB.75.092102}{\bibfield  {journal} 
  {\bibinfo {journal} {Physical Review B}\ }\textbf {\bibinfo {volume} {75}},\ \bibinfo
  {pages} {092102} (\bibinfo {year} {2007})}\BibitemShut {NoStop}%
\bibitem [{\citenamefont {Buchenau}\ and\ \citenamefont
  {Zorn}(1992)}]{Buchenau92}%
  \BibitemOpen
  \bibfield  {author} {\bibinfo {author} {\bibfnamefont {U.}~\bibnamefont
  {Buchenau}}\ and\ \bibinfo {author} {\bibfnamefont {R.}~\bibnamefont
  {Zorn}},\ }\href {http://stacks.iop.org/0295-5075/18/i=6/a=009} {\bibfield
  {journal} {\bibinfo  {journal} {Europhysics Letters}\ }\textbf {\bibinfo
  {volume} {18}},\ \bibinfo {pages} {523} (\bibinfo {year} {1992})}\BibitemShut
  {NoStop}%
\bibitem [{\citenamefont {Sokolov}\ \emph {et~al.}(1993)\citenamefont
  {Sokolov}, \citenamefont {R\"ossler}, \citenamefont {Kisliuk},\ and\
  \citenamefont {Quitmann}}]{Sokolov93}%
  \BibitemOpen
  \bibfield  {author} {\bibinfo {author} {\bibfnamefont {A.~P.}\ \bibnamefont
  {Sokolov}}, \bibinfo {author} {\bibfnamefont {E.}~\bibnamefont {R\"ossler}},
  \bibinfo {author} {\bibfnamefont {A.}~\bibnamefont {Kisliuk}}, \ and\
  \bibinfo {author} {\bibfnamefont {D.}~\bibnamefont {Quitmann}},\ }\href
  {\doibase 10.1103/PhysRevLett.71.2062} {\bibfield  {journal} {\bibinfo
  {journal} {Physical Review Letters}\ }\textbf {\bibinfo {volume} {71}},\
  \bibinfo {pages} {2062} (\bibinfo {year} {1993})}\BibitemShut {NoStop}%
\bibitem [{\citenamefont {Scopigno}\ \emph {et~al.}(2003)\citenamefont
  {Scopigno}, \citenamefont {Ruocco},\ and\ \citenamefont
  {Sette}}]{Scopigno03}%
  \BibitemOpen
  \bibfield  {author} {\bibinfo {author} {\bibfnamefont {T.}~\bibnamefont
  {Scopigno}}, \bibinfo {author} {\bibfnamefont {G.}~\bibnamefont {Ruocco}}, \
  and\ \bibinfo {author} {\bibfnamefont {F.}~\bibnamefont {Sette}},\ }\href
  {\doibase 10.1126/science.1089446} {\bibfield  {journal} {\bibinfo  {journal}
  {Science}\ }\textbf {\bibinfo {volume} {302}},\ \bibinfo {pages} {849}
  (\bibinfo {year} {2003})}\BibitemShut {NoStop}%
\bibitem [{\citenamefont {Novikov}\ and\ \citenamefont
  {Sokolov}(2004)}]{Novikov04}%
  \BibitemOpen
  \bibfield  {author} {\bibinfo {author} {\bibfnamefont {V.~N.}\ \bibnamefont
  {Novikov}}\ and\ \bibinfo {author} {\bibfnamefont {A.~P.}\ \bibnamefont
  {Sokolov}},\ }\href {\doibase 10.1038/nature02947} {\bibfield  {journal}
  {\bibinfo  {journal} {Nature}\ }\textbf {\bibinfo {volume} {431}},\ \bibinfo
  {pages} {961} (\bibinfo {year} {2004})}\BibitemShut {NoStop}%
\bibitem [{\citenamefont {Larini}\ \emph {et~al.}(2008)\citenamefont {Larini},
  \citenamefont {Ottochian}, \citenamefont {de~Michele},\ and\ \citenamefont
  {Leporini}}]{Larini08}%
  \BibitemOpen
  \bibfield  {author} {\bibinfo {author} {\bibfnamefont {L.}~\bibnamefont
  {Larini}}, \bibinfo {author} {\bibfnamefont {A.}~\bibnamefont {Ottochian}},
  \bibinfo {author} {\bibfnamefont {C.}~\bibnamefont {de~Michele}}, \ and\
  \bibinfo {author} {\bibfnamefont {D.}~\bibnamefont {Leporini}},\ }\href
  {\doibase 10.1038/nphys788} {\bibfield  {journal} {\bibinfo  {journal}
  {Nature Physics}\ }\textbf {\bibinfo {volume} {4}},\ \bibinfo {pages} {42}
  (\bibinfo {year} {2008})}\BibitemShut {NoStop}%
\bibitem [{\citenamefont {Ngai}(2004)}]{Ngai04}%
  \BibitemOpen
  \bibfield  {author} {\bibinfo {author} {\bibfnamefont {K.~L.}\ \bibnamefont
  {Ngai}},\ }\href {\doibase 10.1080/14786430310001644080} {\bibfield
  {journal} {\bibinfo  {journal} {Philosophical Magazine}\ }\textbf {\bibinfo
  {volume} {84}},\ \bibinfo {pages} {1341} (\bibinfo {year} {2004})}\BibitemShut {NoStop}%
\bibitem [{\citenamefont {Bernini}\ \emph {et~al.}(2015)\citenamefont
  {Bernini}, \citenamefont {Puosi},\ and\ \citenamefont
  {Leporini}}]{Bernini15}%
  \BibitemOpen
  \bibfield  {author} {\bibinfo {author} {\bibfnamefont {S.}~\bibnamefont
  {Bernini}}, \bibinfo {author} {\bibfnamefont {F.}~\bibnamefont {Puosi}}, \
  and\ \bibinfo {author} {\bibfnamefont {D.}~\bibnamefont {Leporini}},\ }\href
  {\doibase http://dx.doi.org/10.1063/1.4916047} {\bibfield  {journal}
  {\bibinfo  {journal} {Journal of Chemical Physics}\ }\textbf {\bibinfo
  {volume} {142}},\ \bibinfo {eid} {124504} (\bibinfo {year}
  {2015})}\BibitemShut {NoStop}%
\bibitem [{\citenamefont {Yan}\ \emph {et~al.}(2013)\citenamefont {Yan},
  \citenamefont {D{\"u}ring},\ and\ \citenamefont {Wyart}}]{yan13}%
  \BibitemOpen
  \bibfield  {author} {\bibinfo {author} {\bibfnamefont {L.}~\bibnamefont
  {Yan}}, \bibinfo {author} {\bibfnamefont {G.}~\bibnamefont {D{\"u}ring}}, \
  and\ \bibinfo {author} {\bibfnamefont {M.}~\bibnamefont {Wyart}},\ }\href
  {\doibase 10.1073/pnas.1300534110} {\bibfield  {journal} {\bibinfo  {journal}
  {Proceedings of the National Academy of Sciences}\ }\textbf {\bibinfo {volume} {110}},\ \bibinfo {pages} {6307}
  (\bibinfo {year} {2013})}\BibitemShut {NoStop}%
\bibitem [{\citenamefont {Mirigian}\ and\ \citenamefont
  {Schweizer}(2013)}]{mir13}%
  \BibitemOpen
  \bibfield  {author} {\bibinfo {author} {\bibfnamefont {S.}~\bibnamefont
  {Mirigian}}\ and\ \bibinfo {author} {\bibfnamefont {K.~S.}\ \bibnamefont
  {Schweizer}},\ }\href {\doibase 10.1021/jz4018943} {\bibfield  {journal} {\bibinfo  {journal} {Journal of
  Physical Chemistry Letters}\ }\textbf {\bibinfo {volume} {4}},\ \bibinfo {pages}
  {3648} (\bibinfo {year} {2013})}\BibitemShut {NoStop}%
\bibitem [{\citenamefont {Schirmacher}\ \emph {et~al.}(2015)\citenamefont
  {Schirmacher}, \citenamefont {Ruocco},\ and\ \citenamefont
  {Mazzone}}]{sch15}%
  \BibitemOpen
  \bibfield  {author} {\bibinfo {author} {\bibfnamefont {W.}~\bibnamefont
  {Schirmacher}}, \bibinfo {author} {\bibfnamefont {G.}~\bibnamefont {Ruocco}},
  \ and\ \bibinfo {author} {\bibfnamefont {V.}~\bibnamefont {Mazzone}},\ }\href
  {\doibase 10.1103/PhysRevLett.115.015901} {\bibfield  {journal} {\bibinfo
  {journal} {Physical Review Letters}\ }\textbf {\bibinfo {volume} {115}},\ \bibinfo
  {pages} {015901} (\bibinfo {year} {2015})}\BibitemShut {NoStop}%
\bibitem [{\citenamefont {Betancourt}\ \emph {et~al.}(2015)\citenamefont
  {Betancourt}, \citenamefont {Starr},\ and\ \citenamefont {Douglas}}]{bet15}%
  \BibitemOpen
  \bibfield  {author} {\bibinfo {author} {\bibfnamefont {P.~A.}\ \bibnamefont
  {Betancourt}}, \bibinfo {author} {\bibfnamefont {P.~Z. H.~F.}\ \bibnamefont
  {Starr}}, \ and\ \bibinfo {author} {\bibfnamefont {J.~F.}\ \bibnamefont
  {Douglas}},\ }\href {\doibase 10.1073/pnas.1418654112} {\bibfield  {journal}
  {\bibinfo  {journal} {Proceedings of the National Academy of Sciences}\ }\textbf {\bibinfo {volume} {112}},\ \bibinfo
  {pages} {2966} (\bibinfo {year} {2015})}\BibitemShut {NoStop}%
\bibitem [{\citenamefont {Rouxel}(2011)}]{rou11}%
  \BibitemOpen
  \bibfield  {author} {\bibinfo {author} {\bibfnamefont {T.}~\bibnamefont
  {Rouxel}},\ }\href {\doibase http://dx.doi.org/10.1063/1.3656695} {\bibfield
  {journal} {\bibinfo  {journal} {Journal of Chemical Physics}\ }\textbf {\bibinfo {volume}
  {135}},\ \bibinfo {eid} {184501} (\bibinfo {year} {2011})}\BibitemShut
  {NoStop}%
\bibitem [{\citenamefont {Xu}\ and\ \citenamefont {McKenna}(2011)}]{xu11}%
  \BibitemOpen
  \bibfield  {author} {\bibinfo {author} {\bibfnamefont {B.}~\bibnamefont
  {Xu}}\ and\ \bibinfo {author} {\bibfnamefont {G.~B.}\ \bibnamefont
  {McKenna}},\ }\href {\doibase http://dx.doi.org/10.1063/1.3567092} {\bibfield
   {journal} {\bibinfo  {journal} {Journal of Chemical Physics}\ }\textbf {\bibinfo {volume}
  {134}},\ \bibinfo {pages} {124902} (\bibinfo {year} {2011})}\BibitemShut
  {NoStop}%
\bibitem [{\citenamefont {Potuzak}\ \emph {et~al.}(2013)\citenamefont
  {Potuzak}, \citenamefont {Guo}, \citenamefont {Smedskjaer},\ and\
  \citenamefont {Mauro}}]{pot13}%
  \BibitemOpen
  \bibfield  {author} {\bibinfo {author} {\bibfnamefont {M.}~\bibnamefont
  {Potuzak}}, \bibinfo {author} {\bibfnamefont {X.}~\bibnamefont {Guo}},
  \bibinfo {author} {\bibfnamefont {M.~M.}\ \bibnamefont {Smedskjaer}}, \ and\
  \bibinfo {author} {\bibfnamefont {J.~C.}\ \bibnamefont {Mauro}},\ }\href
  {\doibase http://dx.doi.org/10.1063/1.4730525} {\bibfield  {journal}
  {\bibinfo  {journal} {Journal of Chemical Physics}\ }\textbf {\bibinfo {volume} {138}},\
  \bibinfo {eid} {12A501} (\bibinfo {year} {2013})}\BibitemShut {NoStop}%
\bibitem [{\citenamefont {Mirigian}\ and\ \citenamefont
  {Schweizer}(2014)}]{mir14b}%
  \BibitemOpen
  \bibfield  {author} {\bibinfo {author} {\bibfnamefont {S.}~\bibnamefont
  {Mirigian}}\ and\ \bibinfo {author} {\bibfnamefont {K.~S.}\ \bibnamefont
  {Schweizer}},\ }\href {\doibase http://dx.doi.org/10.1063/1.4900507}
  {\bibfield  {journal} {\bibinfo  {journal} {Journal of Chemical Physics}\ }\textbf
  {\bibinfo {volume} {141}},\ \bibinfo {pages} {161103} (\bibinfo {year}
  {2014})}\BibitemShut {NoStop}%
\bibitem [{\citenamefont {Klameth}\ and\ \citenamefont {Vogel}(2015)}]{kla15}%
  \BibitemOpen
  \bibfield  {author} {\bibinfo {author} {\bibfnamefont {F.}~\bibnamefont
  {Klameth}}\ and\ \bibinfo {author} {\bibfnamefont {M.}~\bibnamefont
  {Vogel}},\ }\href@noop {} {\bibfield  {journal} {\bibinfo  {journal}
  {arXiv:1506.05568}\ } (\bibinfo {year} {2015})}\BibitemShut {NoStop}%
\bibitem [{\citenamefont {Krausser}\ \emph {et~al.}(2015)\citenamefont
  {Krausser}, \citenamefont {Samwer},\ and\ \citenamefont {Zaccone}}]{kra15}%
  \BibitemOpen
  \bibfield  {author} {\bibinfo {author} {\bibfnamefont {J.}~\bibnamefont
  {Krausser}}, \bibinfo {author} {\bibfnamefont {K.~H.}\ \bibnamefont
  {Samwer}}, \ and\ \bibinfo {author} {\bibfnamefont {A.}~\bibnamefont
  {Zaccone}},\ }\href {\doibase 10.1073/pnas.1503741112} {\bibfield  {journal}
  {\bibinfo  {journal} {Proceedings of the National Academy of Sciences}\ }\textbf {\bibinfo {volume} {112}},\ \bibinfo
  {pages} {13762} (\bibinfo {year} {2015})}\BibitemShut {NoStop}%
\bibitem [{\citenamefont {Mitrofanov}\ \emph {et~al.}(2016)\citenamefont
  {Mitrofanov}, \citenamefont {Wang}, \citenamefont {Makarov}, \citenamefont
  {Wang},\ and\ \citenamefont {Khonik}}]{mit16}%
  \BibitemOpen
  \bibfield  {author} {\bibinfo {author} {\bibfnamefont {Y.~P.}\ \bibnamefont
  {Mitrofanov}}, \bibinfo {author} {\bibfnamefont {D.~P.}\ \bibnamefont
  {Wang}}, \bibinfo {author} {\bibfnamefont {A.~S.}\ \bibnamefont {Makarov}},
  \bibinfo {author} {\bibfnamefont {W.~H.}\ \bibnamefont {Wang}}, \ and\
  \bibinfo {author} {\bibfnamefont {V.~A.}\ \bibnamefont {Khonik}},\
  }\href {\doibase 10.1038/srep23026} {\bibfield  {journal} {\bibinfo  {journal} {Scientific Reports}\
  }\textbf {\bibinfo {volume} {6}},\ \bibinfo {pages} {23026} (\bibinfo {year}
  {2016})}\BibitemShut {NoStop}%
\bibitem [{\citenamefont {Ikeda}\ and\ \citenamefont {Aniya}(2016)}]{ike16}%
  \BibitemOpen
  \bibfield  {author} {\bibinfo {author} {\bibfnamefont {M.}~\bibnamefont
  {Ikeda}}\ and\ \bibinfo {author} {\bibfnamefont {M.}~\bibnamefont {Aniya}},\
  }\href {\doibase 10.1016/j.jnoncrysol.2015.05.017} {\bibfield  {journal} {\bibinfo  {journal} {Journal of Non-Crystalline Solids}\ }\textbf {\bibinfo {volume} {431}},\ \bibinfo {pages} {52} (\bibinfo
  {year} {2016})}\BibitemShut {NoStop}%
\bibitem [{\citenamefont {Syutkin}(2013)}]{syu13}%
  \BibitemOpen
  \bibfield  {author} {\bibinfo {author} {\bibfnamefont {V.~M.}\ \bibnamefont
  {Syutkin}},\ }\href {\doibase http://dx.doi.org/10.1063/1.4821752} {\bibfield
   {journal} {\bibinfo  {journal} {Journal of Chemical Physics}\ }\textbf {\bibinfo {volume}
  {139}},\ \bibinfo {pages} {114506} (\bibinfo {year} {2013})}\BibitemShut
  {NoStop}%
\bibitem [{\citenamefont {Liu}(2015)}]{Liu15}%
  \BibitemOpen
  \bibfield  {author} {\bibinfo {author} {\bibfnamefont {W.}\ \bibnamefont
  {Liu}},\ and\ \bibinfo {author} {\bibfnamefont {L.}~\bibnamefont {Zhang}},\
  }\href {\doibase http://dx.doi.org/10.1364/AO.54.006841} {\bibfield
   {journal} {\bibinfo  {journal} {Applied Optics}\ }\textbf {\bibinfo {volume}
  {54}},\ \bibinfo {pages} {6841} (\bibinfo {year} {2015})}\BibitemShut
  {NoStop}%
\bibitem [{\citenamefont {Jakobsen}\ \emph {et~al.}(2011)\citenamefont
  {Jakobsen}, \citenamefont {Niss}, \citenamefont {Maggi}, \citenamefont
  {Olsen}, \citenamefont {Christensen},\ and\ \citenamefont
  {Dyre}}]{Jakobsen11}%
  \BibitemOpen
  \bibfield  {author} {\bibinfo {author} {\bibfnamefont {B.}~\bibnamefont
  {Jakobsen}}, \bibinfo {author} {\bibfnamefont {K.}~\bibnamefont {Niss}},
  \bibinfo {author} {\bibfnamefont {C.}~\bibnamefont {Maggi}}, \bibinfo
  {author} {\bibfnamefont {N.~B.}\ \bibnamefont {Olsen}}, \bibinfo {author}
  {\bibfnamefont {T.}~\bibnamefont {Christensen}}, \ and\ \bibinfo {author}
  {\bibfnamefont {J.~C.}\ \bibnamefont {Dyre}},\ }\href {\doibase
  10.1016/j.jnoncrysol.2010.08.010} {\bibfield  {journal} {\bibinfo  {journal}
  {Journal of Non-Crystalline Solids}\ }\textbf {\bibinfo {volume} {357}},\
  \bibinfo {pages} {267} (\bibinfo {year} {2011})}\BibitemShut {NoStop}%
\bibitem [{\citenamefont {Niss}\ \emph {et~al.}(2010)\citenamefont {Niss},
  \citenamefont {Dalle-Ferrier}, \citenamefont {Frick}, \citenamefont {Russo},
  \citenamefont {Dyre.},\ and\ \citenamefont {Alba-Simionesco}}]{Niss10}%
  \BibitemOpen
  \bibfield  {author} {\bibinfo {author} {\bibfnamefont {K.}~\bibnamefont
  {Niss}}, \bibinfo {author} {\bibfnamefont {C.}~\bibnamefont {Dalle-Ferrier}},
  \bibinfo {author} {\bibfnamefont {B.}~\bibnamefont {Frick}}, \bibinfo
  {author} {\bibfnamefont {D.}~\bibnamefont {Russo}}, \bibinfo {author}
  {\bibfnamefont {J.}~\bibnamefont {Dyre}}, \ and\ \bibinfo {author}
  {\bibfnamefont {C.}~\bibnamefont {Alba-Simionesco}},\ }\href {\doibase
  10.1103/PhysRevE.82.021508} {\bibfield  {journal} {\bibinfo  {journal}
  {Physical Review E}\
  }\textbf {\bibinfo {volume} {82}},\ \bibinfo {pages} {021508} (\bibinfo
  {year} {2010})}\BibitemShut {NoStop}%
\bibitem [{\citenamefont {Buchenau}\ \emph {et~al.}(2014)\citenamefont
  {Buchenau}, \citenamefont {Zorn},\ and\ \citenamefont {Ramos}}]{Buchenau14}%
  \BibitemOpen
  \bibfield  {author} {\bibinfo {author} {\bibfnamefont {U.}~\bibnamefont
  {Buchenau}}, \bibinfo {author} {\bibfnamefont {R.}~\bibnamefont {Zorn}}, \
  and\ \bibinfo {author} {\bibfnamefont {M.~A.}\ \bibnamefont {Ramos}},\ }\href
  {\doibase 10.1103/PhysRevE.90.042312} {\bibfield  {journal} {\bibinfo
  {journal} {Physical Review E}\ }\textbf {\bibinfo {volume} {90}},\ \bibinfo
  {pages} {042312} (\bibinfo {year} {2014})}\BibitemShut {NoStop}%
\bibitem [{\citenamefont {Klieber}\ \emph {et~al.}(2013)\citenamefont
  {Klieber}, \citenamefont {Hecksher}, \citenamefont {Pezeril}, \citenamefont
  {Torchinsky}, \citenamefont {Dyre},\ and\ \citenamefont
  {Nelson}}]{Klieber13}%
  \BibitemOpen
  \bibfield  {author} {\bibinfo {author} {\bibfnamefont {C.}~\bibnamefont
  {Klieber}}, \bibinfo {author} {\bibfnamefont {T.}~\bibnamefont {Hecksher}},
  \bibinfo {author} {\bibfnamefont {T.}~\bibnamefont {Pezeril}}, \bibinfo
  {author} {\bibfnamefont {D.~H.}\ \bibnamefont {Torchinsky}}, \bibinfo
  {author} {\bibfnamefont {J.}~\bibnamefont {Dyre}}, \ and\ \bibinfo {author}
  {\bibfnamefont {K.~A.}\ \bibnamefont {Nelson}},\ }\href {\doibase
  10.1063/1.4789948} {\bibfield  {journal} {\bibinfo  {journal} {Journal of
  Chemical Physics}\ }\textbf {\bibinfo {volume} {138}},\ \bibinfo {pages}
  {12A544} (\bibinfo {year} {2013})}\BibitemShut {NoStop}%
\bibitem [{\citenamefont {Hecksher}\ and\ \citenamefont
  {Dyre}(2015)}]{Hecksher15}%
  \BibitemOpen
  \bibfield  {author} {\bibinfo {author} {\bibfnamefont {T.}~\bibnamefont
  {Hecksher}}\ and\ \bibinfo {author} {\bibfnamefont {J.~C.}\ \bibnamefont
  {Dyre}},\ }\href {\doibase 10.1016/j.jnoncrysol.2014.08.056} {\bibfield
  {journal} {\bibinfo  {journal} {Journal of Non-Crystalline Solids}\ }\textbf
  {\bibinfo {volume} {407}},\ \bibinfo {pages} {14} (\bibinfo {year}
  {2015})}\BibitemShut {NoStop}%
\bibitem [{\citenamefont {Gundermann}(2013)}]{Gundermann13_thesis}%
  \BibitemOpen
  \bibfield  {author} {\bibinfo {author} {\bibfnamefont {D.}~\bibnamefont
  {Gundermann}},\ }\href
  {http://glass.ruc.dk/pdf/phd_afhandlinger/ditte_thesis.pdf} {\enquote
  {\bibinfo {title} {Testing predictions of the isomorph theory by
  experiment},}\ } (\bibinfo {year} {2013}),\ \bibinfo {note} {Ph{D} thesis,
  {R}oskilde {U}niversity, {DNRF} centre ``{G}lass \& {T}ime''}\BibitemShut
  {NoStop}%
\bibitem [{\citenamefont {Xiao}\ \emph {et~al.}(2015)\citenamefont {Xiao},
  \citenamefont {Tofteskov}, \citenamefont {Christensen}, \citenamefont
  {Dyre},\ and\ \citenamefont {Niss}}]{Xiao15}%
  \BibitemOpen
  \bibfield  {author} {\bibinfo {author} {\bibfnamefont {W.}~\bibnamefont
  {Xiao}}, \bibinfo {author} {\bibfnamefont {J.}~\bibnamefont {Tofteskov}},
  \bibinfo {author} {\bibfnamefont {T.~V.}\ \bibnamefont {Christensen}},
  \bibinfo {author} {\bibfnamefont {J.~C.}\ \bibnamefont {Dyre}}, \ and\
  \bibinfo {author} {\bibfnamefont {K.}~\bibnamefont {Niss}},\ }\href {\doibase
  10.1016/j.jnoncrysol.2014.08.041} {\bibfield  {journal} {\bibinfo  {journal}
  {Journal of Non-Crystalline Solids}\ }\textbf {\bibinfo {volume} {407}},\
  \bibinfo {pages} {190} (\bibinfo {year} {2015})}\BibitemShut {NoStop}%
\bibitem [{\citenamefont {Hansen}()}]{Hansen16}%
  \BibitemOpen
  \bibfield  {author} {\bibinfo {author} {\bibfnamefont {H.~W.}\ \bibnamefont
  {Hansen}},\ }\href@noop {} {\enquote {\bibinfo {title} {(unpublished)},}\
  }\BibitemShut {NoStop}%
\bibitem [{\citenamefont {Jakobsen}\ \emph {et~al.}(2005)\citenamefont
  {Jakobsen}, \citenamefont {Niss},\ and\ \citenamefont {Olsen}}]{Jakobsen05}%
  \BibitemOpen
  \bibfield  {author} {\bibinfo {author} {\bibfnamefont {B.}~\bibnamefont
  {Jakobsen}}, \bibinfo {author} {\bibfnamefont {K.}~\bibnamefont {Niss}}, \
  and\ \bibinfo {author} {\bibfnamefont {N.~B.}\ \bibnamefont {Olsen}},\ }\href
  {\doibase 10.1063/1.2136887} {\bibfield  {journal} {\bibinfo  {journal} {Journal of Chemical Physics}\ }\textbf {\bibinfo {volume} {123}},\ \bibinfo
  {pages} {234511} (\bibinfo {year} {2005})}\BibitemShut {NoStop}%
\bibitem [{\citenamefont {Niss}(2007)}]{Niss07_thesis}%
  \BibitemOpen
  \bibfield  {author} {\bibinfo {author} {\bibfnamefont {K.}~\bibnamefont
  {Niss}},\ }\href {http://dirac.ruc.dk/~kniss/KristineNissPhDthesis.pdf}
  {\enquote {\bibinfo {title} {Fast and slow dynamics of glass-forming liquids
  -- {W}hat can we learn from high pressure experiment?}}\ } (\bibinfo {year}
  {2007}),\ \bibinfo {note} {PhD thesis}\BibitemShut {NoStop}%
\bibitem [{\citenamefont {Dyre}(1998)}]{Dyre98}%
  \BibitemOpen
  \bibfield  {author} {\bibinfo {author} {\bibfnamefont {J.~C.}\ \bibnamefont
  {Dyre}},\ }\href {\doibase 10.1016/S0022-3093(98)00502-X} {\bibfield  {journal} {\bibinfo  {journal} {Journal
  of Non-Crystalline Solids}\ }\textbf {\bibinfo {volume} {235-237}},\ \bibinfo
  {pages} {142} (\bibinfo {year} {1998})}\BibitemShut {NoStop}%
\bibitem [{\citenamefont {Frick}\ \emph {et~al.}(2012)\citenamefont {Frick},
  \citenamefont {Combet},\ and\ \citenamefont {van Eijck}}]{Frick12}%
  \BibitemOpen
  \bibfield  {author} {\bibinfo {author} {\bibfnamefont {B.}~\bibnamefont
  {Frick}}, \bibinfo {author} {\bibfnamefont {J.}~\bibnamefont {Combet}}, \
  and\ \bibinfo {author} {\bibfnamefont {L.}~\bibnamefont {van Eijck}},\ }\href
  {\doibase 10.1016/j.nima.2011.11.090} {\bibfield  {journal} {\bibinfo
  {journal} {Nuclear Instruments and Methods in Physics Research A}\ }\textbf
  {\bibinfo {volume} {669}},\ \bibinfo {pages} {7} (\bibinfo {year}
  {2012})}\BibitemShut {NoStop}%
\bibitem [{\citenamefont {Bridgman}(1949)}]{Bridgman49}%
  \BibitemOpen
  \bibfield  {author} {\bibinfo {author} {\bibfnamefont {P.~W.}\ \bibnamefont
  {Bridgman}},\ }\href {\doibase 10.2307/20023533} {\bibfield  {journal}
  {\bibinfo  {journal} {Proceedings of the American Academy of Arts and
  Sciences}\ }\textbf {\bibinfo {volume} {77}},\ \bibinfo {pages} {129}
  (\bibinfo {year} {1949})}\BibitemShut {NoStop}%
\bibitem [{\citenamefont {Barlow}\ \emph {et~al.}(1966)\citenamefont {Barlow},
  \citenamefont {Lamb},\ and\ \citenamefont {Matheson}}]{Barlow66}%
  \BibitemOpen
  \bibfield  {author} {\bibinfo {author} {\bibfnamefont {A.~J.}\ \bibnamefont
  {Barlow}}, \bibinfo {author} {\bibfnamefont {J.}~\bibnamefont {Lamb}}, \ and\
  \bibinfo {author} {\bibfnamefont {A.~J.}\ \bibnamefont {Matheson}},\
  }\href {\doibase 10.1098/rspa.1966.0138 } {\bibfield  {journal} {\bibinfo  {journal} {Proceedings of the
  Royal Society of London. Series A, Mathematical, Physical and Engineering Sciences}\
  }\textbf {\bibinfo {volume} {292}},\ \bibinfo {pages} {322} (\bibinfo {year}
  {1966})}\BibitemShut {NoStop}%
\bibitem [{\citenamefont {Li}\ \emph {et~al.}(1995)\citenamefont {Li},
  \citenamefont {H.~E.~King}, \citenamefont {Oliver}, \citenamefont {Herbst},\
  and\ \citenamefont {Cummins}}]{Li95}%
  \BibitemOpen
  \bibfield  {author} {\bibinfo {author} {\bibfnamefont {G.}~\bibnamefont
  {Li}}, \bibinfo {author} {\bibfnamefont {H.~E.}~\bibnamefont {King}},
  \bibinfo {author} {\bibfnamefont {W.~F.}\ \bibnamefont {Oliver}}, \bibinfo
  {author} {\bibfnamefont {C.~A.}\ \bibnamefont {Herbst}}, \ and\ \bibinfo
  {author} {\bibfnamefont {H.~Z.}\ \bibnamefont {Cummins}},\ }\href {\doibase 10.1103/PhysRevLett.74.2280}
  {\bibfield  {journal} {\bibinfo  {journal} {Physical Review Letters}\
  }\textbf {\bibinfo {volume} {74}},\ \bibinfo {pages} {2280} (\bibinfo {year}
  {1995})}\BibitemShut {NoStop}%
\bibitem [{\citenamefont {J.~Sekine}\ and\ \citenamefont
  {Mine}(2006)}]{Masahara06}%
  \BibitemOpen
  \bibfield  {author} {\bibinfo {author} {\bibfnamefont {M.~Oguni},\ \bibnamefont
  {J.~Sekine}}\ and\ \bibinfo {author} {\bibfnamefont {H.}~\bibnamefont
  {Mine}},\ }\href {\doibase 10.1016/j.jnoncrysol.2006.02.169} {\bibfield  {journal} {\bibinfo  {journal} {Journal
  of Non-Crystalline Solids}\ }\textbf {\bibinfo {volume} {352}},\ \bibinfo
  {pages} {4665} (\bibinfo {year} {2006})}\BibitemShut {NoStop}%
\bibitem [{\citenamefont {Roed}\ \emph {et~al.}(2015)\citenamefont {Roed},
  \citenamefont {Niss},\ and\ \citenamefont {Jakobsen}}]{Roed15}%
  \BibitemOpen
  \bibfield  {author} {\bibinfo {author} {\bibfnamefont {L.~A.}\ \bibnamefont
  {Roed}}, \bibinfo {author} {\bibfnamefont {K.}~\bibnamefont {Niss}}, \ and\
  \bibinfo {author} {\bibfnamefont {B.}~\bibnamefont {Jakobsen}},\ }\href
  {\doibase 10.1063/1.4936867} {\bibfield  {journal} {\bibinfo  {journal} {Journal of Chemical Physics}\ }\textbf {\bibinfo {volume} {143}},\ \bibinfo
  {pages} {221101} (\bibinfo {year} {2015})}\BibitemShut {NoStop}%
\bibitem [{\citenamefont {Rahman}\ \emph {et~al.}(1962)\citenamefont {Rahman},
  \citenamefont {Singwi},\ and\ \citenamefont {Sj\"olander}}]{Rahman62}%
  \BibitemOpen
  \bibfield  {author} {\bibinfo {author} {\bibfnamefont {A.}~\bibnamefont
  {Rahman}}, \bibinfo {author} {\bibfnamefont {K.}~\bibnamefont {Singwi}}, \
  and\ \bibinfo {author} {\bibfnamefont {A.}~\bibnamefont {Sj\"olander}},\
  }\href {\doibase http://dx.doi.org/10.1103/PhysRev.126.986} {\bibfield
  {journal} {\bibinfo  {journal} {Physical Review}\ }\textbf {\bibinfo {volume}
  {126}},\ \bibinfo {pages} {986} (\bibinfo {year} {1962})}\BibitemShut
  {NoStop}%
\bibitem [{\citenamefont {Christensen}\ and\ \citenamefont
  {Olsen}(1995)}]{Christensen95}%
  \BibitemOpen
  \bibfield  {author} {\bibinfo {author} {\bibfnamefont {T.}~\bibnamefont
  {Christensen}}\ and\ \bibinfo {author} {\bibfnamefont {N.}~\bibnamefont
  {Olsen}},\ }\href {\doibase 10.1063/1.1146126} {\bibfield  {journal}
  {\bibinfo  {journal} {Review of Scientific Instruments}\ }\textbf {\bibinfo
  {volume} {66}},\ \bibinfo {pages} {5019} (\bibinfo {year}
  {1995})}\BibitemShut {NoStop}%
\bibitem [{\citenamefont {Dyre}\ and\ \citenamefont {Wang}(2012)}]{Dyre12}%
  \BibitemOpen
  \bibfield  {author} {\bibinfo {author} {\bibfnamefont {J.~C.}\ \bibnamefont
  {Dyre}}\ and\ \bibinfo {author} {\bibfnamefont {W.~H.}\ \bibnamefont
  {Wang}},\ }\href {\doibase
  10.1063/1.4724102}{\bibfield  {journal} {\bibinfo  {journal} {Journal of Chemical Physics}\ }\textbf {\bibinfo {volume} {136}},\ \bibinfo
  {pages} {224108} (\bibinfo {year} {2012})}\BibitemShut {NoStop}%
\bibitem [{\citenamefont {Barlow}\ \emph {et~al.}(1967)\citenamefont {Barlow},
  \citenamefont {Erginsav},\ and\ \citenamefont {Lamb}}]{Barlow67}%
  \BibitemOpen
  \bibfield  {author} {\bibinfo {author} {\bibfnamefont {A.~J.}\ \bibnamefont
  {Barlow}}, \bibinfo {author} {\bibfnamefont {A.}~\bibnamefont {Erginsav}}, \
  and\ \bibinfo {author} {\bibfnamefont {J.}~\bibnamefont {Lamb}},\ }\href
  {\doibase 10.1098/rspa.1967.0116} {\bibfield  {journal} {\bibinfo  {journal}
  {Proceedings of the Royal Society A}\ }\textbf {\bibinfo {volume} {298}},\
  \bibinfo {pages} {481} (\bibinfo {year} {1967})}\BibitemShut {NoStop}%
\bibitem [{\citenamefont {Harrison}(1976)}]{Harrison76}%
  \BibitemOpen
  \bibfield  {author} {\bibinfo {author} {\bibfnamefont {G.}~\bibnamefont
  {Harrison}},\ }\href@noop {} {\emph {\bibinfo {title} {The Dynamic Properties
  of Supercooled Liquids}}}\ (\bibinfo  {publisher} {Academic Press, New
  York},\ \bibinfo {year} {1976})\BibitemShut {NoStop}%
\bibitem [{\citenamefont {Capaccioli}\ \emph {et~al.}(2012)\citenamefont
  {Capaccioli}, \citenamefont {Ngai}, \citenamefont {Ancherbak},\ and\
  \citenamefont {Paciaroni}}]{Capaccioli12}%
  \BibitemOpen
  \bibfield  {author} {\bibinfo {author} {\bibfnamefont {S.}~\bibnamefont
  {Capaccioli}}, \bibinfo {author} {\bibfnamefont {K.~L.}\ \bibnamefont
  {Ngai}}, \bibinfo {author} {\bibfnamefont {S.}~\bibnamefont {Ancherbak}}, \
  and\ \bibinfo {author} {\bibfnamefont {A.}~\bibnamefont {Paciaroni}},\ }\href
  {\doibase 10.1021/jp2057892} {\bibfield  {journal} {\bibinfo  {journal} {Journal of Chemical Physics}\ }\textbf {\bibinfo {volume} {116}},\ \bibinfo
  {pages} {1745} (\bibinfo {year} {2012})}\BibitemShut {NoStop}%
\bibitem [{\citenamefont {Hecksher}\ \emph {et~al.}(2013)\citenamefont
  {Hecksher}, \citenamefont {Olsen}, \citenamefont {Nelson},\ and\
  \citenamefont {Dyre}}]{Hecksher13}%
  \BibitemOpen
  \bibfield  {author} {\bibinfo {author} {\bibfnamefont {T.}~\bibnamefont
  {Hecksher}}, \bibinfo {author} {\bibfnamefont {N.~B.}\ \bibnamefont {Olsen}},
  \bibinfo {author} {\bibfnamefont {K.~A.}\ \bibnamefont {Nelson}}, \ and\
  \bibinfo {author} {\bibfnamefont {J.~C.}\ \bibnamefont {Dyre}},\ }\href
  {\doibase 10.1063/1.4789946} {\bibfield  {journal} {\bibinfo  {journal}
  {Journal of Physical Chemistry}\ }\textbf {\bibinfo {volume} {138}},\
  \bibinfo {pages} {12A543} (\bibinfo {year} {2013})}\BibitemShut {NoStop}%
\bibitem [{\citenamefont {Roed}\ \emph {et~al.}(2013)\citenamefont {Roed},
  \citenamefont {Gundermann}, \citenamefont {Dyre},\ and\ \citenamefont
  {Niss}}]{Roed13}%
  \BibitemOpen
  \bibfield  {author} {\bibinfo {author} {\bibfnamefont {L.}~\bibnamefont
  {Roed}}, \bibinfo {author} {\bibfnamefont {D.}~\bibnamefont {Gundermann}},
  \bibinfo {author} {\bibfnamefont {J.~C.}\ \bibnamefont {Dyre}}, \ and\
  \bibinfo {author} {\bibfnamefont {K.}~\bibnamefont {Niss}},\ }\href {\doibase
  10.1063/1.4821163} {\bibfield  {journal} {\bibinfo  {journal} {Journal of
  Chemical Physics}\ }\textbf {\bibinfo {volume} {139}},\ \bibinfo {pages}
  {101101} (\bibinfo {year} {2013})}\BibitemShut {NoStop}%
\bibitem [{\citenamefont {Frick}\ \emph {et~al.}(1988)\citenamefont {Frick},
  \citenamefont {Richter}, \citenamefont {Petry},\ and\ \citenamefont
  {Buchenau}}]{Frick88}%
  \BibitemOpen
  \bibfield  {author} {\bibinfo {author} {\bibfnamefont {B.}~\bibnamefont
  {Frick}}, \bibinfo {author} {\bibfnamefont {D.}~\bibnamefont {Richter}},
  \bibinfo {author} {\bibfnamefont {W.}\ \bibnamefont {Petry}}, \ and\
  \bibinfo {author} {\bibfnamefont {U.}~\bibnamefont {Buchenau}},\ }\href {\doibase
 10.1007/BF01320541} {\bibfield  {journal} {\bibinfo  {journal} {Zeitschrift für Physik B Condensed Matter}\ }\textbf {\bibinfo {volume} {70}},\ \bibinfo {pages}
  {73--79} (\bibinfo {year} {1988})}\BibitemShut {NoStop}%

\end{thebibliography}

%

\end{document}